\documentclass{article}

\usepackage{arxiv}

\usepackage[utf8]{inputenc} 
\usepackage[T1]{fontenc}    
\usepackage{lmodern}        
\usepackage{hyperref}       
\usepackage{url}            
\usepackage{booktabs}       
\usepackage{amsfonts}       
\usepackage{nicefrac}       
\usepackage{microtype}      
\usepackage{graphicx}

\title{Estimating default probability and correlation using Stan}

\author{
    Jesus A. Pinera-Esquivel
   \\
     \\
   \\
  \texttt{} \\
  }

\usepackage{longtable,booktabs,array}
\usepackage{calc} 
\usepackage{etoolbox}
\makeatletter
\patchcmd\longtable{\par}{\if@noskipsec\mbox{}\fi\par}{}{}
\makeatother
\IfFileExists{footnotehyper.sty}{\usepackage{footnotehyper}}{\usepackage{footnote}}
\makesavenoteenv{longtable}

\newlength{\cslhangindent}
\newlength{\csllabelwidth}
\newlength{\cslentryspacingunit} 
  {}%
  {\par}
\newenvironment{CSLReferences}[2] 
 {
  \setlength{\parindent}{0pt}
  \ifodd #1
  \let\oldpar\par
  \def\par{\hangindent=\cslhangindent\oldpar}
  \fi
  \setlength{\parskip}{#2\cslentryspacingunit}
 }%
 {}
\usepackage{calc}

\usepackage{amsthm, amsmath,amsfonts}
\usepackage{bm}
\usepackage{float}
\usepackage{pgfplots}
\usepackage[boxed]{algorithm2e}

\counterwithin*{theo_aux}{section}

\pgfplotsset{compat=1.18}

\makeatletter
\newcommand*{\Fbar}{}%
\DeclareRobustCommand*{\Fbar}{%
  \mathpalette\@Fbar{}%
}
\newcommand*{\@Fbar}[2]{%
  \sbox0{$#1\mathrm{F}\m@th$}%
  \sbox2{$#1F\m@th$}%
  \rlap{%
    \hbox to\wd2{%
      \hfill
      $\overline{%
        \vrule width 0pt height\ht0 %
        \kern\wd0 %
      }$%
    }%
  }%
  \copy2 %
}
\makeatother
\begin{document}
\maketitle

\begin{abstract}
This work has the objective of estimating default probabilities and correlations of credit portfolios given default rate information through a Bayesian framework using Stan. We use Vasicek's single factor credit model to establish the theoretical framework for the behavior of the default rates, and use NUTS Markov Chain Monte Carlo to estimate the parameters.

We compare the Bayesian estimates with classical estimates such as moments estimators and maximum likelihood estimates. We apply the methodology both to simulated data and to corporate default rates, and perform inferences through Bayesian methods in order to exhibit the advantages of such a framework. We perform default forecasting and exhibit the importance of an adequate estimation of default correlations, and exhibit the advantage of using Stan to perform sampling regarding prior choice.
\end{abstract}


\newpage

\hypertarget{introduction}{%
\section{Introduction}\label{introduction}}

There is a wide variety of tools regarding the measurement of credit risk. For example, there are both structural and reduced form models, both requiring different types of information\footnote{For a deeper explanation of the differences, see Jarrow and Protter (2004).}. In this text, we focus in the Vasicek single factor credit model (Vasicek, 2002), which can be thought as a hybrid model, and the use of Bayesian techniques to calibrate such a model given only information regarding the number of defaults and outstanding debtors.

This text is thought of as a continuation of the work of Pfeuffer, Nagl, Fischer, and Rösch (2020), as a way of comparing classic moments and likelihood estimations with Bayesian estimation using Stan\footnote{We use the implementation of \textbf{CmdStanR} developed by Gabry and Češnovar (2022).}. The analysis relies heavily on the package \textbf{AssetCorr} developed by Nagl et al. (2021) to estimate the correlation through different methods, and the package \textbf{CmdStanR} developed by Gabry and Češnovar (2022) to estimate the correlation through Bayesian methods.

The document is constructed as follows. In the first part, we introduce the Vasicek model and how it can be used to both estimate default correlations and the whole loss distribution of a given credit portfolio. After, we mention different estimation methods that are used as benchmarks, which include both moment and likelihood estimators. We also mention the construction of the Bayesian model. In the third part, we perform inferences using data from S\&P's Annual Global Corporate Default report, which contains annual information of defaults from 1981 to 2021. We compare the estimates of the classical methods with the Bayesian framework, both point estimates and intervals. We also perform model checking and appropriate convergence of the Bayesian estimates. Finally, we mention some concluding remarks regarding the possible improvement of using Bayesian methods for this type of models, and we suggest new avenues of study.

\hypertarget{literature-review}{%
\subsection{Literature review}\label{literature-review}}

There is a wide variety of studies regarding the estimation of factor credit models using Bayesian methods. Tasche (2013) assumes uniform priors for the default probability and finds explicit bounds for default probabilities assuming both independent and dependent defaults. Chang and Yu (2014) uses a Bayesian approach along the generalized method of moments and the Gibbs sampler to estimate credibility intervals for default probability and correlation assuming auto-correlated latent factors. Kazianka (2016) develops what is called objective priors through the use of reference priors for both the Gaussian and \(t\) single factor models to estimate de probability of default. Finally, Blümke (2020) estimates no observed defaults probabilities of default using sequential Bayesian updating.

Regarding more complex models, Bu (2014) uses generalized linear mixed models and survival analysis to model to model defaults using latent factors, using Bayesian inference throughout. Castro (2012) uses non-linear non-Gaussian state space models and Gibbs sampler to estimate a multi-factor dynamic model, grouping defaults in buckets and assuming a dynamic conditional default probability with VAR(1) factors. Park, Sirakaya, and Kim (2010) use dynamic hierarchical Bayesian models to analyze defaults in Korean companies between 2000 and 2003.

We are interested in studying Bayesian estimation through the framework explained in Gabry, Simpson, Vehtari, Betancourt, and Gelman (2019), focusing in the use of Stan language to perform the inference of the model. We focus in the simple case of the Gaussian single-factor model to assess the ability of the Bayesian workflow to correctly infer the relevant parameters, comparing it to classical methods, following Pfeuffer et al. (2020).

\hypertarget{the-vasicek-factor-credit-model}{%
\subsection{The Vasicek factor credit model}\label{the-vasicek-factor-credit-model}}

The factor model has its roots on Merton's option model, developed by Vasicek (1987) and refined later by himself (Vasicek, 2002). Even though the model was initially deduced through the use of stochastic calculus, Vasicek (2002) developed the same results through the usage of normal random variables. As before, let there be a credit portfolio with \(N\) debtors, each represented by a Bernoulli r.v. \(I_n \sim Bernoulli(p_n)\). The underlying assumption is that
\[
I_n = \pmb{1}_{\{X_n\leq u_n\}},\quad n = 1,...,N
\]
where \(X_n\) is called a \textbf{latent variable}, which is non-observable\footnote{Usually, this variable represents the assets of the debtor, but it is more closely related to its creditworthiness in a general scenario.} and \(u_n\) is a threshold below which the credit is considered defaulted. For simplicity, Vasicek (2002) assumes that \(p_n = p\) for all \(n=1,...,N\) and that the credit balances are all equal to \(f\). Moreover, the latent variables are assumed of the form
\[
X_n = \sqrt{\rho}Z + \sqrt{1-\rho}\; \varepsilon_n
\]
where \(Z\) and \(\varepsilon_n\) are i.i.d. random variables distributed \(N(0,1)\). One can observe that \(I_n\) are conditionally independent given the factor \(Z\) (solving for \(\varepsilon_n\) gives the result). In fact, the total defaults are such that
\[
D|Z = \sum_{n=1}^N I_n |Z = Bin(N,\pi(Z))
\]
where
\begin{equation}\label{eq:pi-cond}
\pi(Z) = \left( \frac{\Phi^{-1}(p)-\sqrt{\rho}Z}{\sqrt{1-\rho}} \right).
\end{equation}
We call \(\pi(Z)\) the conditional default probability, which has a density of the form
\[
f_{\pi}(x) = \sqrt{\frac{1-\rho}{\rho}}\exp\left( -\frac{1}{2\rho}(\sqrt{1-\rho} \, \Phi^{-1}(x)- \Phi^{-1}(p))^2 + \frac{1}{2} (\Phi^{-1}(x))^2 \right).
\]
Thus, we say \(\pi \sim Vas(p,\rho)\) if it has the above density.

The following figures show this density for different values of \(\rho\) and \(p\). One can see that both parameters play an important role in the shape and location of the density\footnote{The higher the correlation parameter, the more mass at the endpoints of 0 and 1. The probability parameter both displaces the distribution and makes it more skewed depending on the value.}. Moreover, one can achieve a wide range of forms for the density such as right or left skewed depending on the parameters.

\begin{figure}[!ht]

{\centering \includegraphics{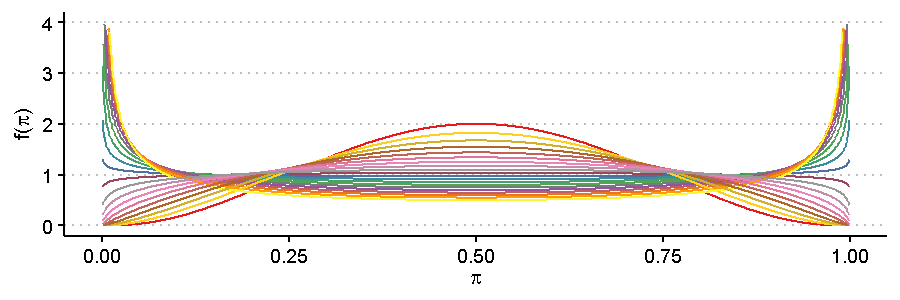} 

}

\caption{Vasicek density for different values of the default correlation, with fixed default probability in 0.5}\label{fig:fig1-1}
\end{figure}

\begin{figure}[!ht]

{\centering \includegraphics{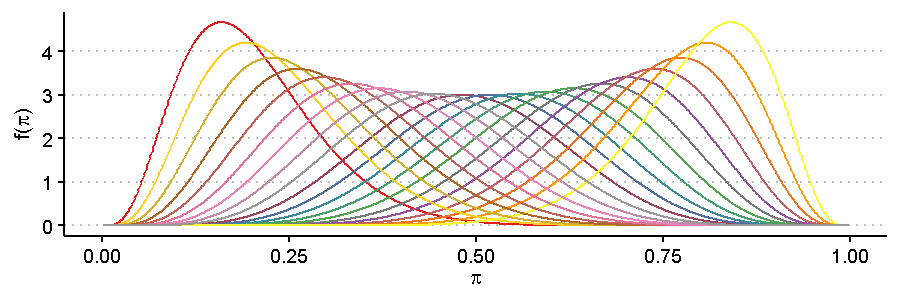} 

}

\caption{Vasicek density for different values of the default probability, with fixed default correlation in 0.1}\label{fig:fig1-2}
\end{figure}

It is worth mentioning that there exists a multi-factor Vasicek model, which will not be studied in this document Its definition is left for the Appendix and its estimation through Bayesian tools and Stan is left for further studies.

\newpage

\hypertarget{estimation-methods-for-the-vasicek-model}{%
\section{Estimation methods for the Vasicek model}\label{estimation-methods-for-the-vasicek-model}}

The above model permits the estimation of the different default probabilities and correlations through different methods. In this section, we first mention some standard estimation methods for these parameters, following the work of Pfeuffer et al. (2020). Next, we present the Bayesian framework to produce estimates of the parameters, and perform mixing and predictive checks in order to asses if the model is appropriate.

\hypertarget{classical-methods}{%
\subsection{Classical methods}\label{classical-methods}}

The most widely used method for parameter estimation is maximum likelihood estimation (MLE). Given our model, we have that \(D_t|\pi_t \sim Bin(N_t,\pi_t)\) and \(\pi_t\sim Vas(p,\rho)\). Note also that \(\pi_t = \pi(Z_t)\) where \(Z_t\sim N(0,1)\), as established in \eqref{eq:pi-cond}. Thus, to apply maximum likelihood, we need to find \(f(D_t)\) for each \(t = 1,...,n\) with \(n\) the length of the series. To this end, we can use the fact that \(D_t|Z_t \sim Bin(N_t,\pi(Z_t))\) instead, such that
\[
f(D_t;p,\rho) = \int_{\mathbb{R}}\binom{N_t}{D_t} \pi(z;p,\rho)^{D_t}(1-\pi(z;p,\rho))^{N_t-D_t}\phi(z)dz
\]
where \(\pi(z;p,\rho) = \Phi((\Phi^{-1}(p)-\sqrt{\rho}z)/\sqrt{1-\rho})\) and \(\phi(z)\) is the standard normal density.

The problem is, then
\[
\max_{p,\rho}L(p,\rho)= \prod_{t=1}^n f(D_t;p,\rho).
\]
Of course, \(f(D_t;p,\rho)\) does not have a closed form, so numerical methods most be used to approximate it. Note that, if the portfolio is too large, \(N_t\to \infty\), numerical integration could fail due to the factor \((1-\pi(z;p,\rho))\), which tends to zero as \(N_t\) grows. Caution must be taken then, taking log-probabilities instead, or using asymptotic relations.

Of course, this method is not the only one available to estimate the parameters of the Vasicek model. As mentioned, for very large portfolios, Gordy (2000) shows that the expectation and variance of the default rates tend to the values of the conditional probability. One can show that \(E[\pi_t] = p\) and that
\[
\text{Var}(\pi_t) = \Phi_2(\Phi^{-1}(p),\Phi^{-1}(p);\rho) - p^2.
\]
Assuming that \(N_t \to\infty\) we have that \(E[D_t/N_t]\to p\) and \(\text{Var}(D_t/N_t)\to \Phi_2(\Phi^{-1}(p),\Phi^{-1}(p);\rho) - p^2\). Thus, given estimates of the expected default rate and the variance of the default rates (through sample mean and variance, for instance), one can estimate the parameters of the model through moment matching.

Other methods of estimation include the corrected moment estimator of Frei and Wunsch (2018), considering an adjustment assuming that the default rates follow a first order autorregresive process, the Beta Value at Risk (VaR) matching of Botha and Vuuren (2010), and the Joint Default Probability Matching estimator of De Servigny and Renault (2002). All of this methods, among others, can be used through the R package \textbf{AssetCorr}.

\hypertarget{bayesian-framework}{%
\subsection{Bayesian framework}\label{bayesian-framework}}

First, we focus on the single factor model. As mentioned before, we assume that the defaults behave conditionally independent given the latent, or conditional, probability of default, such that \(D_t | \pi_t \sim Bin(N_t,\pi_t)\). The conditional probability of default is given by \(\pi_t|p,\rho\sim Vas(p,\rho)\) such that \(\pi_t\) is as in \eqref{eq:pi-cond}.

Before, we regarded \(p\) and \(\rho\) as numeric parameters to be estimated. However, now we introduce them as random variables to perform Bayesian inference. As such, we need to establish prior distributions for the parameters. Throughout this section, and when speaking in Bayesian terms, we avoid the usage of subscripts in densities and distributions, and allow the variable to determine which distribution we are referring to. For example, instead of using \(F_\theta(x)\) we use \(F(\theta)\).

First, we construct the full posterior distribution to properly sample from it. Given a series of default time series \(D_t\) and credits \(N_t\) with \(t = 1,...,n\), we assume that \(D_t|\pi_t \sim Bin(N_t,\pi_t)\), and that \(\pi_t|p,\rho \sim Vas(p,\rho)\). Denoting \(D = (D_1,...,D_n)\), \(N = (N_1,...,N_n)\), and \(\pi = (\pi_1,...,\pi_n)\), we have the posterior distribution
\[
f(\pi,p,\rho|D,N) \propto f(D|\pi,p,\rho,N)f(\pi,p,\rho).
\]

Given that \(D\) depends only on \(\pi\) and \(N\), we have that
\[
f(D|\pi,p,\rho,N) =f(D|\pi,N) = \prod_{t=1}^n Bin(N_t,\pi_t)
\]
where \(Bin(n,p)\) denotes a binomial probability mass function. Moreover, note that
\[
\begin{aligned}
f(\pi,p,\rho) &= f(\pi|p,\rho)f(p,\rho)\\
&=\left(\prod_{t=1}^n Vas(p,\rho)\right) f(p,\rho).
\end{aligned}
\]

Thus, the posterior can be expressed as
\[
f(\pi,p,\rho|D,N) \propto \left(\prod_{t=1}^n Bin(N_t,\pi_t)\right)\left(\prod_{t=1}^n Vas(p,\rho)\right) f(p,\rho).
\]

To choose appropriate prior distributions for the parameters \(p\) and \(\rho\), we take into account the knowledge of default portfolios. Firstly, it is clear that the probability of default must lie between 0 and 1, so \(p\in(0,1)\). Moreover, the model requires that \(\rho >0\), so we also have that \(\rho\in (0,1)\). Thus, we must have priors that reflect such bounds

However, we can better identify the priors using our knowledge regarding the model and the nature of credit portfolios. For example, we rarely expect the probability to pass 50\%. In fact, the default rate being more than 25\% already indicates a very risky credit portfolio. As such, we suggest a prior for \(p\) such that most, though not all, of its mass is before \(1/2\), with a mode around \(0.25\), the middle point. For instance, using a beta proportion distribution\footnote{The \(BetaP(\mu,\phi)\) distribution, or beta proportion distribution, is a reparametrization of the beta distribution \(Beta(\alpha,\beta)\) such that \(\alpha = \mu\phi\) and \(\beta = (1-\mu)\phi\), giving a constant mean \(\mu\) and variance inversely related to \(\phi\).} with mean \(\mu_p = 1/5\) and a dispersion parameter small enough so the prior is not as informative.

Regarding the correlation, note that if \(\pi \sim Vas(p,\rho)\), then \(\rho\) appears to be a variance parameter. Indeed, if we observe the simulated default series, a high default correlation implies a more volatile default series. Thus, it might make sense to relate the \(p\) parameter to the correlation parameter through \(f(p,\rho) = f(p|\rho)f(\rho)\) such that \(f(p|\rho)\) is related to \(\rho\) through a variance parameter. For instance, using \(p\sim BetaP(\mu,a\rho)\) where \(a\) is a positive number.

Therefore, we define the joint prior density \(f(p,\rho) = f(p|\rho) f(\rho)\) where \(f(\rho)\) is \(BetaP(\mu_\rho,\phi_\rho)\) and \(f(p|\rho) = BetaP(\mu_p,a\rho)\) with \(a>0\). In the Appendix we show a prior sensitivity analysis for the parameters \(\mu_\rho\), \(\phi_\rho\), \(\mu_p\) and \(a\) taking simulated examples. However, for some of these parameters, their standard values can be established in order to have weakly informative priors: \(\mu_\rho=1/2\) relates to our lack of knowledge of the value of the default correlation, though we expect it to be almost uniform around \(1/2\) the middle point between 0 and 1; \(\mu_p=1/5\) relates to our belief that most default probabilities are below 0.5 and likely around 0.25 but with a right skew. To perform sampling, we use the library \textbf{CmdStanR}, which implements Markov Chain Monte Carlo through the No U-Turn sampler (NUTS)\footnote{The efficiency of the NUTS sampler relative to the Gibbs and Hamiltonian samplers is still under study. Its main advantage lies in performing as well as an appropriately tuned Hamilton sampler without the need of tuning the steps' parameters (Hoffman and Gelman, 2014). Studies such as Al Hakmani and Sheng (2019) and Nishio and Arakawa (2019) show that, for certain applications, the NUTS sampler performs better than the Gibbs sampler.}.

\hypertarget{estimation-of-parameters-with-simulated-data}{%
\subsection{Estimation of parameters with simulated data}\label{estimation-of-parameters-with-simulated-data}}

As a first example, we take simulated time series with different combinations of default probabilities and correlations. We take 4 time series, each of size 20.

\begin{table}[!h]

\caption{\label{tab:unnamed-chunk-2}Time series parameters for simulated data}
\centering
\begin{tabular}[t]{lcc}
\toprule
  & Probability & Correlation\\
\midrule
Low probability and correlation (LL) & 0.01 & 0.1\\
Low probability, high correlation (LH) & 0.01 & 0.5\\
Hig probability, low correlation (HL) & 0.05 & 0.1\\
High probability and correlation (HH) & 0.05 & 0.5\\
\bottomrule
\end{tabular}
\end{table}

\begin{figure}[!ht]

{\centering \includegraphics{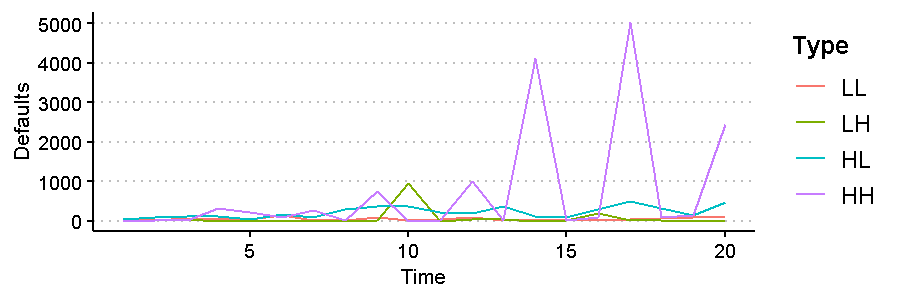} 

}

\caption{Default series for the different types of probabilities and correlations}\label{fig:fig2-1}
\end{figure}

\noindent We use the package \textbf{AssetCor} to simulate the series. Figure \ref{fig:fig2-1} shows the simulated time series of the defaults for 20 points of data\footnote{Usually, default data is yearly, but there are instances of monthly data, so we do not give a specific time dimension.} where we assume that the number of credits is of the form \(N_t = at+b+e_t\) where \(e_t\sim N(0,\sigma_N)\). For these series we have \(a = 500\), \(b = 1000\) and \(\sigma_N = 500\). Note how the low correlation defaults have smaller variance, which is expected. As the correlation tends to 1, the defaults behave more like only on debtor, so if one debtor defaults, many do at the same time.

First, we estimate the correlation parameter using the asymptotic MLE method (AMLE) and the corrected method of moments (CMM), detailed in Gordy (2000) and Frei and Wunsch (2018) respectively using the \textbf{AssetCorr} implementation. We also produce 10 thousand bootstrap replicates to correct possible biases and also produce 95\% bootstrap confidence intervals\footnote{By default, the bias-corrected and accelerated intervals are used.}.

\begin{figure}[!ht]

{\centering \includegraphics{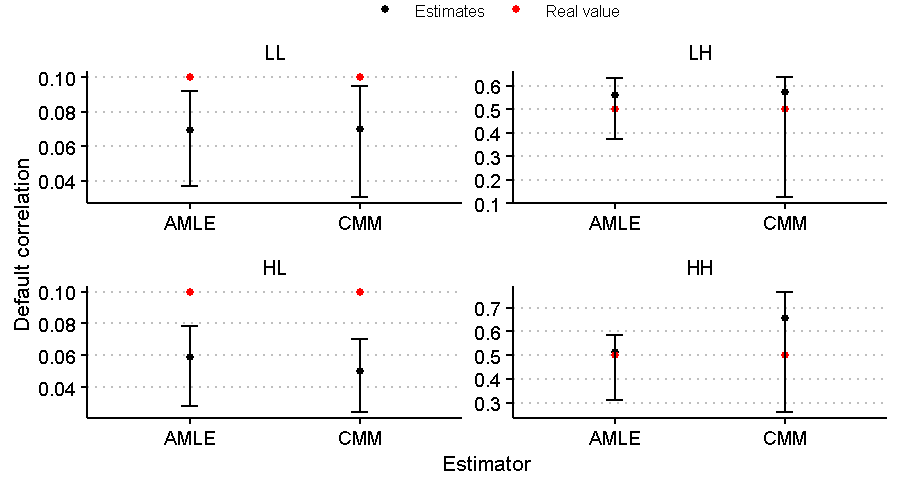} 

}

\caption{Point estimates and 95\% bootstrap confidence intervals for the default correlation for each default series through AMLE and CMM methods}\label{fig:fig2-2}
\end{figure}

Figure \ref{fig:fig2-2} shows graphs of the estimate and intervals for each series and each type of estimate. We observe that two of the intervals contain the real value of the default correlation, with the low correlation intervals not capturing the real values\footnote{This is primary due to the bias induced by the use of the AMLE and CMM method, which require \(N\) large or a long time series (Pfeuffer et al., 2020). In fact, the AMLE requires both \(N\) large and a long time series. In the appendix we provide a cumulative estimate to observe how the estimate behaves as the sample size increases for the AMLE method.}.

\begin{figure}[!ht]

{\centering \includegraphics{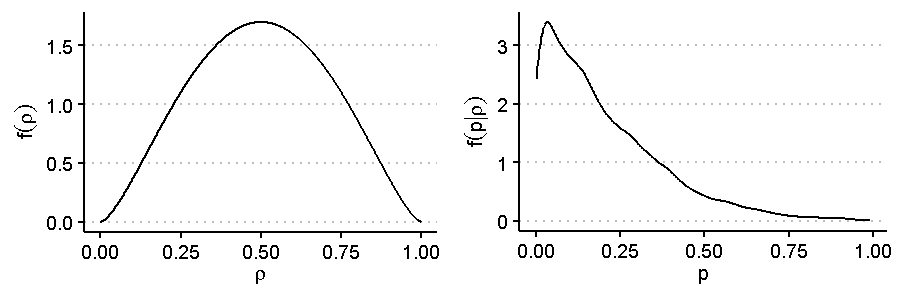} 

}

\caption{Prior densities for $\rho$ and $p$ assuming Beta proportion priors $f(p|\rho)$ and $f(\rho)$}\label{fig:fig2-3}
\end{figure}

Now, we perform inferences for each series using a Bayesian approach. For these inferences we use the prior parameters values \(\mu_\rho = 0.5\), \(\mu_p = 0.2\), \(\phi = 5\) and \(a = 10\). Figure \ref{fig:fig2-3} shows the priors for both parameters. We note that \(f(\rho)\) has a bell-shaped density, with most of its mass concentrated around 0.5, though with positive probability throughout the interval. We produce the density of \(p|\rho\) through sampling of \(\rho \sim f(\rho)\) and \(p\sim f(p|\rho)\). We observe a distribution with most of its mass between 0 and 0.5, though with a heavy tail covering the whole interval. We believe these are appropriate priors for the task at hand.

\begin{figure}[!ht]

{\centering \includegraphics{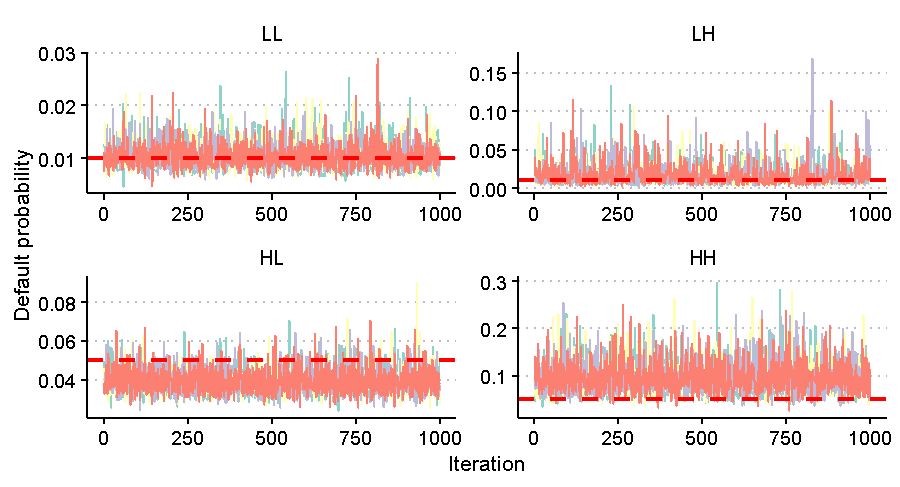} 

}

\caption{Chains for $p$ parameter for each type of series, with the real value in red}\label{fig:fig2-4}
\end{figure}

Using the above priors and the posterior distribution, we perform the sampling through the NUTS implementation of MCMC in Stan using 4 chains. The plots in Figure \ref{fig:fig2-4} show the chains for each parameter of the model and each type of series (low or high probability or correlation). All the chains appear to have mixed correctly, and the real parameter is contained within the chains.

\begin{figure}[!ht]

{\centering \includegraphics{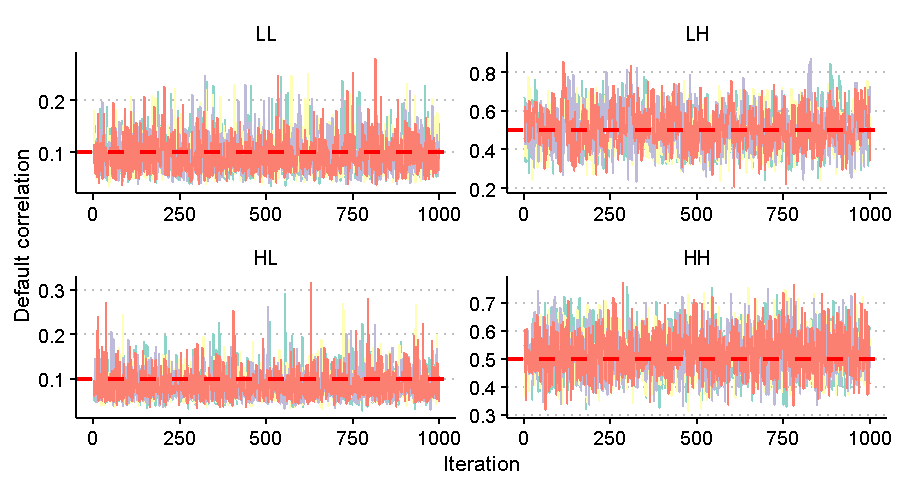} 

}

\caption{Chains for $\rho$ parameter for each type of series, with the real value in red}\label{fig:fig2-5}
\end{figure}

We observe something similar with the correlation parameter, as shown in Figure \ref{fig:fig2-5}. The correct mixing of the chains is also assessed through \(\hat{R}\) values below 1.01 for all parameters, with an Effective Sample Size (ESS) greater than 1 thousand (around 3 thousand for most cases) for all parameters.

\begin{figure}[!ht]

{\centering \includegraphics{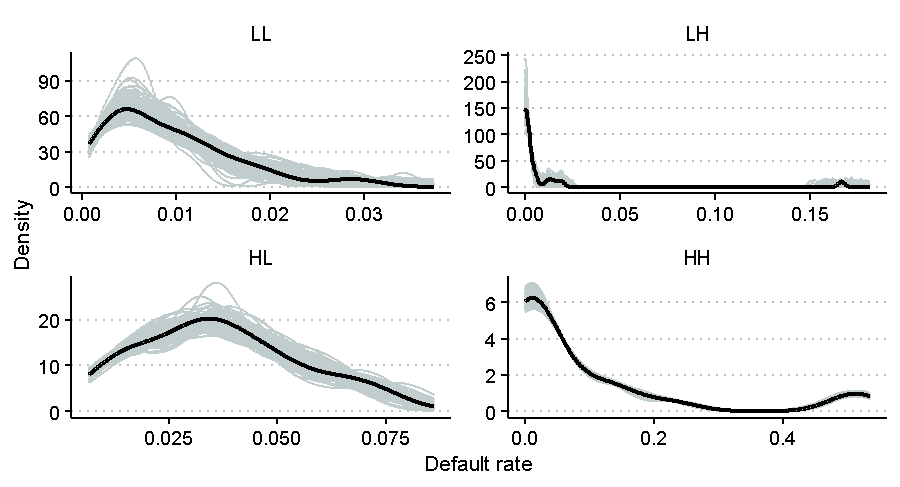} 

}

\caption{Density overlay of the posterior distribution through 500 draws and the observed distribution}\label{fig:fig2-6}
\end{figure}

Regarding the appropriate fit to the data, Figure \ref{fig:fig2-6} plots a density overlay of the posterior predictive distribution \(f(\tilde{D}|\pi,p,\rho)\) for each series. We observe adequate fits for the empirical densities in all the series, though the LH probability behavior is hard to capture. We can also observe how the correlation parameter impacts the distribution of defaults. The higher the correlation, the higher the variance of the distribution. Note, for instance, how for a low probability of 1\% there are times in which rates reach almost 20\%.

\begin{figure}[!ht]

{\centering \includegraphics{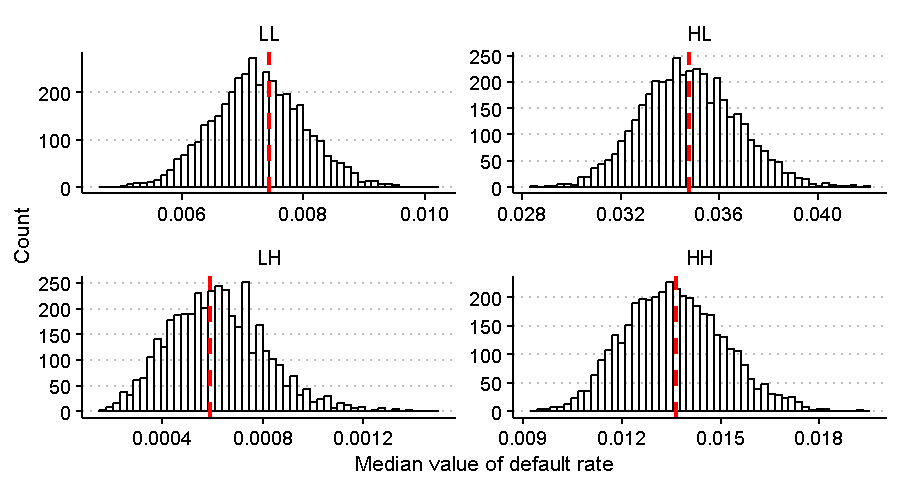} 

}

\caption{Median value of the posterior draws, and the median of the observed series in red}\label{fig:fig2-7}
\end{figure}

Moreover, following with the appropriateness of the model, we also calculate statistics for the median and the range\footnote{The range is defined as the difference between the third and first quartiles.} of the defaults, and compare them to the observed values in the simulated data. In Figure \ref{fig:fig2-7} we observe that the sample median lies around the middle of the posterior draws values, with a p-value of approximately 0.5 for all the series\footnote{Given a statistic function \(T\) of the sample \(D_{obs}\), such as the median, and posterior draws of the sample \(D^{(1)},D^{(2)},..., D^{(S)}\), each of the same size as \(D_{obs}\), we loosely define the p-value of model \(M\) as the probability \(\mathbb{P}(T(D_{obs})\leq T(D)|M)\), which we estimate with \(\sum_{k=1}^S \pmb{1}\{T(D_{obs})\leq T(D^{(k)}) \}/S\). The better the fit of the model, the closer the p-value is to 0.5. In a sense, is a way of measuring how well the model fits the data.}. Something similar occurs with the range, although for the low probability-high correlation case we have a less than optimal fit, though still adequate with a p-value of approximately 0.25.

Finally, we compare the bootstrap confidence intervals of the two previous methods with the 95\% credible interval for the correlation parameter, with the point the mean value. In Figure \ref{fig:fig2-9} we can see the previous intervals with the Bayesian estimator. We observe not only that the credible interval contains the real value in all cases, but the mean point estimate is highly precise, with the highest absolute relative difference being in the HL case of 12\%.

\begin{figure}[!ht]

{\centering \includegraphics{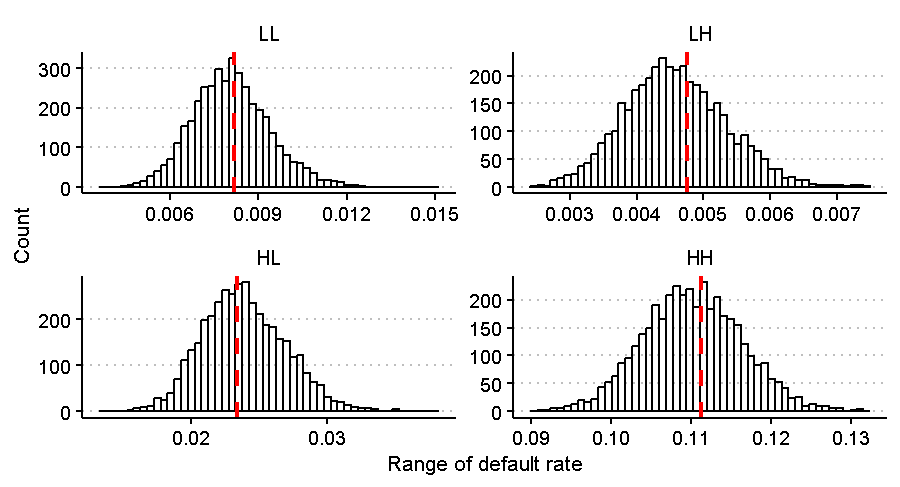} 

}

\caption{Range of the posterior draws, and the range of the observed series in red}\label{fig:fig2-8}
\end{figure}

\begin{figure}[!htb]

{\centering \includegraphics{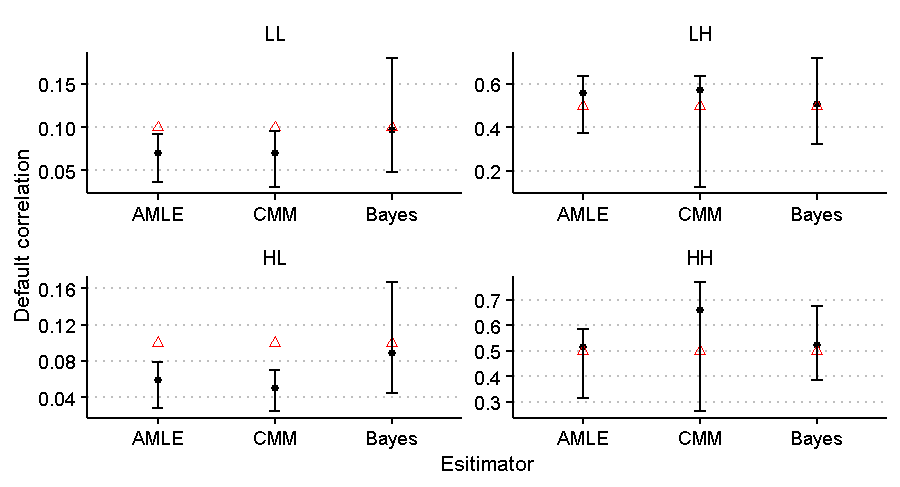} 

}

\caption{Confidence intervals at 95\% for classical methods and 95\% credibility interval, with the mean as the middle point}\label{fig:fig2-9}
\end{figure}

\newpage

\hypertarget{analysis-of-sp-corporate-defaults}{%
\section{Analysis of S\&P Corporate Defaults}\label{analysis-of-sp-corporate-defaults}}

In this section, we analyze the default rates of corporate debt published by S\&P Global Ratings, separating investment grade (IG) and speculative grade (SG) defaults. The data is publicly available in the \textbf{2021 Annual Global Corporate Default And Rating Transition Study}.

\hypertarget{exploratory-analysis}{%
\subsection{Exploratory analysis}\label{exploratory-analysis}}

\begin{figure}[!ht]

{\centering \includegraphics{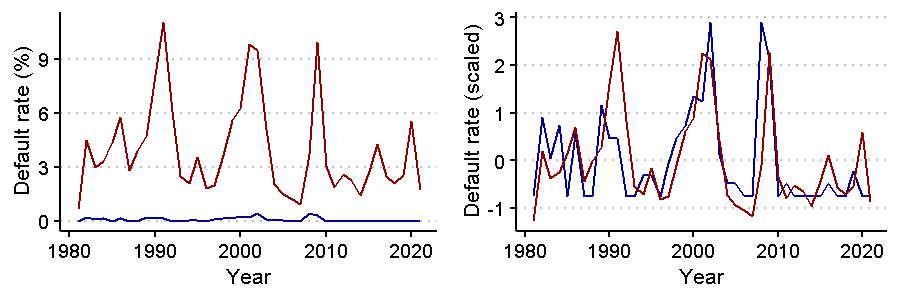} 

}

\caption{Default rates (in percentage) and scaled default rates (scaled by historical mean and standard deviation) of investment (blue) and speculative grade firms (red)}\label{fig:fig3-1}
\end{figure}

First, we analyze the relevant time series and their characteristics. We have a total of 41 default rates for each group. Figure \ref{fig:fig3-1} shows the annual default rates for each group. We can observe, first, that the default rate of SG corporations is much higher than for IG. The figure also shows the scaled default rate, in order to asses differences in variability. We can observe that variability seems to be similar.

\begin{figure}[!ht]

{\centering \includegraphics{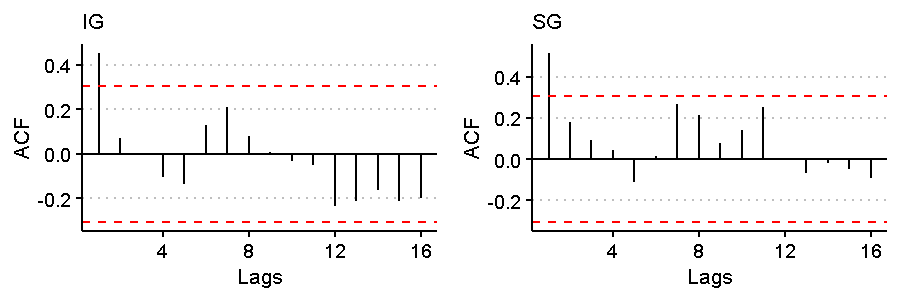} 

}

\caption{Autocorrelation function for the defaults for each type of firm}\label{fig:fig3-2}
\end{figure}

Regarding the behavior of the default time series, we can observe in Figure \ref{fig:fig3-2} the autocorrelation function of the defaults. We can clearly observe a first order autocorrelation, which agrees with the findings of Frei and Wunsch (2018), and the further adjustment of the method of moments to incorporate AR behaviors in the defaults.

\hypertarget{estimation-of-default-probability-and-correlation-for-corporate-debt}{%
\subsection{Estimation of default probability and correlation for corporate debt}\label{estimation-of-default-probability-and-correlation-for-corporate-debt}}

\begin{figure}[!ht]

{\centering \includegraphics{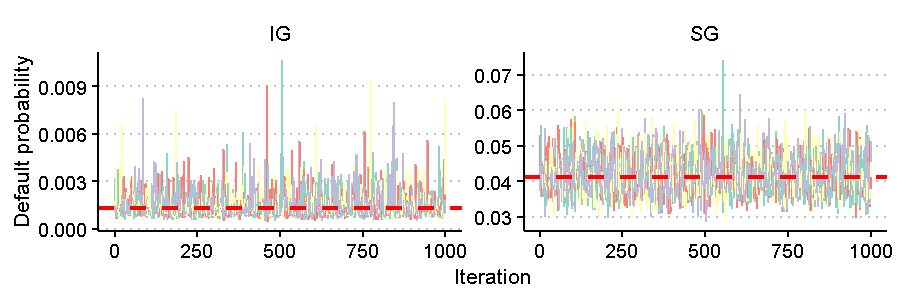} 

}

\caption{Sampled values for the default probability for different chains, with the median value in red}\label{fig:fig3-3}
\end{figure}

As with the simulated data, we perform inference through NUTS MCMC in Stan for the default probability and correlation, for both IG and SG firms. We use the same prior parameters as in the simulation exercise, but change \(\mu_p = 0.1\) given smaller default rates. Figure \ref{fig:fig3-3} shows the chains\footnote{All parameters have \(\hat{R}\) values below 1.01.} for the default probability, for each type of firm. We can observe that the default probability of IG firms is much smaller than the SG default probability. In fact, the estimated probability of the event \(\mathbb{P}\{p_{I}>p_S\}\) is 0.

\begin{figure}[!ht]

{\centering \includegraphics{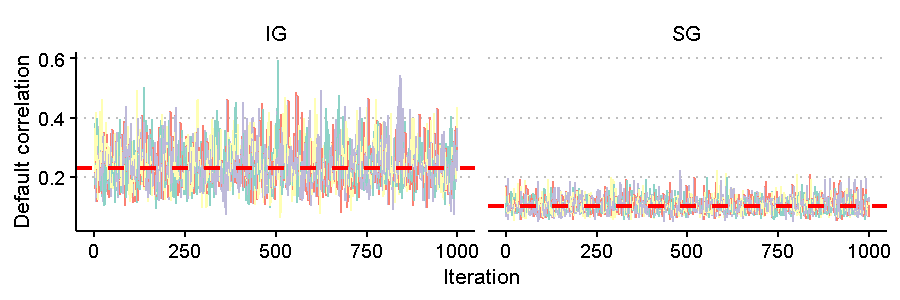} 

}

\caption{Sampled values for the default correlation for different chains, with the median value in red and in same scale}\label{fig:fig3-4}
\end{figure}

Regarding the correlation, Figure \ref{fig:fig3-4} shows the sampled chains, which exhibit good mixing as with the default probability. Moreover, unlike with the probability, the correlation of IG firms is higher than the SG correlation. The probability \(\mathbb{P}\{\rho_{I}>\rho_S\}\) is estimated at around 0.99. This might indicate that, even though IG firms have much smaller default probability, their tail risk might be higher due to a higher default correlation and thus a higher default variance.

\begin{figure}[!ht]

{\centering \includegraphics{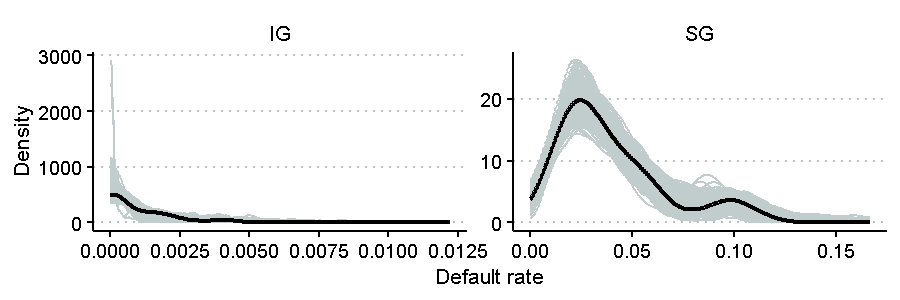} 

}

\caption{Sampled values for the default correlation for different chains, with the median value in red, with same scale}\label{fig:fig3-5}
\end{figure}

\begin{figure}[!ht]

{\centering \includegraphics{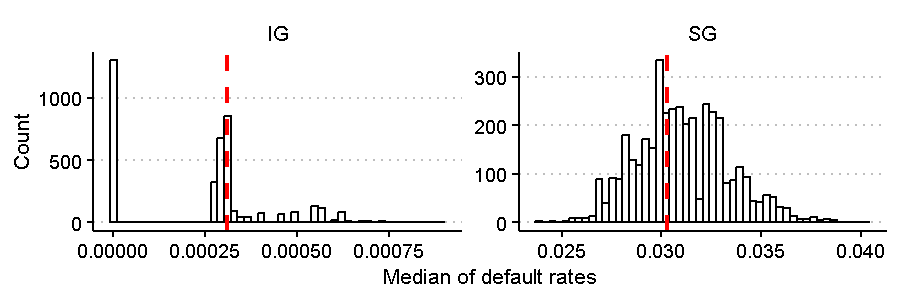} 

}

\caption{Sampled median default rate values for each chain and the observed median in red, for each type of firm}\label{fig:fig3-6}
\end{figure}

To assess model adequacy, we plot a density overlay with 500 samples. Figure \ref{fig:fig3-5} shows the observed density and the sampled density. We observe that the densities overlay, but, unlike the simulation exercises, the sampled densities are more variable. The median value, shown in Figure \ref{fig:fig3-6}, is also adequate, with p-values above 20\%, or below 80\%, with the observed value landing approximately near the median of the samples. It is worth noting that the fit for the IG firms, shows median values of zero, given the very low default probability and high default correlation.

Finally, as a concluding analysis of the S\&P data, we perform predictive analysis to assess if the model can accurately predict future defaults. We will perform a cumulative fitting procedure, starting with 10 defaults and sequentially adding more observations, performing a one step forecast using the number of credits of the following period. This is, we perform inferences using data up until year \(t = 10,11,...,40\) and predict \(D_{t+1}\) given \(N_{t+1}\), which are the number of credits at the start of the year.

\begin{figure}[!ht]

{\centering \includegraphics{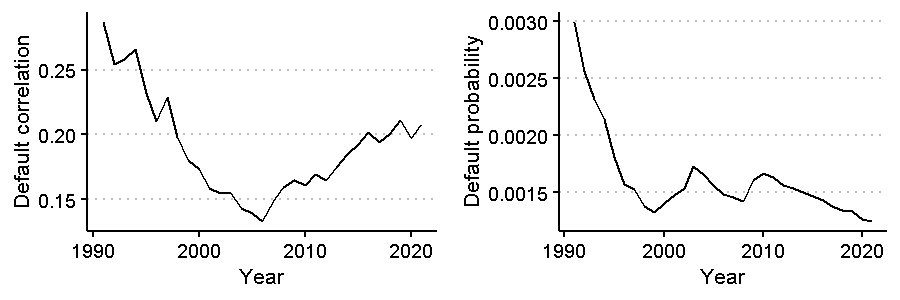} 

}

\caption{Estimated default probability and correlation through time using up-to-date data for IG firms}\label{fig:fig3-7}
\end{figure}

\begin{figure}[!ht]

{\centering \includegraphics{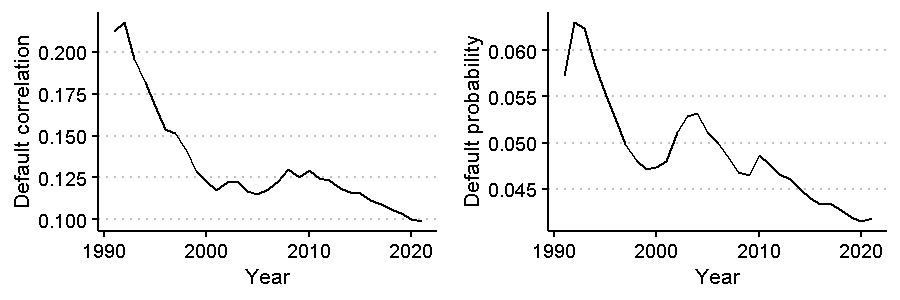} 

}

\caption{Estimated default probability and correlation through time using up-to-date data for SG firms}\label{fig:fig3-8}
\end{figure}

Figures \ref{fig:fig3-7} and \ref{fig:fig3-8} show the median value of the sampled default probability and correlation for each type of firm, IG and SG respectively, through the increasing sample size procedure. We can observe that all the values appear to decrease over time, probably due to a larger sample size, giving more precise inferences. In the case of IG firms, the correlation ``jumped backed'' approximately in the great financial crisis.

\begin{figure}[!ht]

{\centering \includegraphics{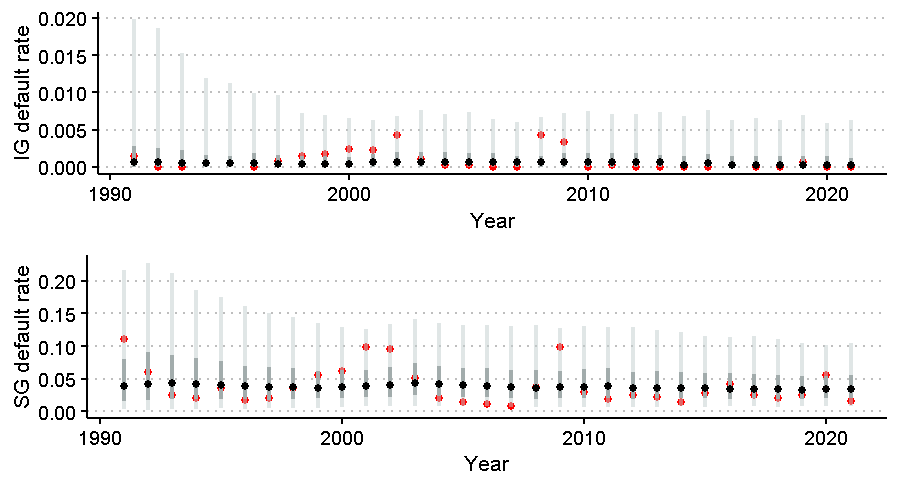} 

}

\caption{50\% and 90\% credibility intervals, and median value of the default rate one-step ahead forecast, with the realized defaul rate in red}\label{fig:fig3-9}
\end{figure}

Regarding the forecasts, Figure \ref{fig:fig3-9} plots the 50\% and 90\% credibility intervals, along the median value, for the one-step default rate \(D_{t+1}|D_1,...,D_t\), with the realized default rate in red. We can observe that for both types of firms, the realized default rates land inside the 90\% interval, even during times of high financial stress such as the DotCom bubble and the Great Financial Crisis. One can also observe how the higher IG correlation impacts the intervals, which are much wider relative to the median value.

\hypertarget{conclusions-and-future-studies}{%
\section{Conclusions and future studies}\label{conclusions-and-future-studies}}

This document shows how the Vasicek model can be applied to model default rates, and how inferences can be made through Stan's NUTS sampler. The results clearly show that the model generates an appropriate description of the default rates process, and that it can accurately capture the parameters of simulated data. Moreover, the Stan implementation allows for general definitions for the priors, allowing us to define more complex models and provide inferences in a short amount of time. Finally, all the architecture surrounding Stan allows us also to easily perform posterior predictive checking, and readily assess appropriate sampling.

The model, however, is still limited, for it does not take into account the observed autocorrelation of defaults. Moreover, multivariate models should be used in order to estimate both intra correlation and inter (between groups) correlation, probably using hierarchical modelling. Thus, multivariate time series models might be appropriate to model defaults in a more complete manner.

\hypertarget{references}{%
\section*{References}\label{references}}
\addcontentsline{toc}{section}{References}

\setlength{\parindent}{0cm}

\hypertarget{refs}{}
\begin{CSLReferences}{1}{0}
\leavevmode\vadjust pre{\hypertarget{ref-bu_2014}{}}%
Bu, Y. (2014). \emph{Bayesian analysis of default and credit migration: Latent factor models for event count and time-to-event data} (PhD thesis). Heriot-Watt University.

\leavevmode\vadjust pre{\hypertarget{ref-al_hakmani_2019}{}}%
Al Hakmani, R., and Sheng, Y. (2019). NUTS for mixture IRT models. In M. Wiberg, S. Culpepper, R. Janssen, J. Gonzalez, \& D. Molenaar (Eds.), \emph{Quantitative psychology} (pp. 25--37). Cham: Springer International Publishing.

\leavevmode\vadjust pre{\hypertarget{ref-blumke_2020}{}}%
Blümke, O. (2020). Estimating the probability of default for no-default and low-default portfolios. \emph{Journal of the Royal Statistical Society Series C: Applied Statistics}, \emph{69}(1), 89--107.

\leavevmode\vadjust pre{\hypertarget{ref-botha_2010}{}}%
Botha, M., and Vuuren, G. van. (2010). Implied asset correlation in retail loan portfolios. \emph{Journal of Risk Management in Financial Institutions}, \emph{3}(2), 156--173.

\leavevmode\vadjust pre{\hypertarget{ref-castro_2012}{}}%
Castro, C. (2012). Confidence sets for asset correlations in portfolio credit risk. \emph{Revista de Economia Del Rosario}, \emph{15}(1), 19--58.

\leavevmode\vadjust pre{\hypertarget{ref-chang_2014}{}}%
Chang, Y.-P., and Yu, C.-T. (2014). Bayesian confidence intervals for probability of default and asset correlation of portfolio credit risk. \emph{Computational Statistics}, \emph{29}, 331--361.

\leavevmode\vadjust pre{\hypertarget{ref-de_servigny_2002}{}}%
De Servigny, A., and Renault, O. (2002). Default correlation: Empirical evidence. \emph{Standard and Poor's}.

\leavevmode\vadjust pre{\hypertarget{ref-frei_2018}{}}%
Frei, C., and Wunsch, M. (2018). Moment estimators for autocorrelated time series and their application to default correlations. \emph{Journal of Credit Risk}, \emph{14}(1), 1--29.

\leavevmode\vadjust pre{\hypertarget{ref-gabry_2022}{}}%
Gabry, J., and Češnovar, R. (2022). \emph{Cmdstanr: R interface to 'CmdStan'}.

\leavevmode\vadjust pre{\hypertarget{ref-gabry_2019}{}}%
Gabry, J., Simpson, D., Vehtari, A., Betancourt, M., and Gelman, A. (2019). Visualization in bayesian workflow. \emph{Journal of the Royal Statistical Society Series A: Statistics in Society}, \emph{182}(2), 389--402.

\leavevmode\vadjust pre{\hypertarget{ref-gordy_2000}{}}%
Gordy, M. B. (2000). A comparative anatomy of credit risk models. \emph{Journal of Banking \& Finance}, \emph{24}(1-2), 119--149.

\leavevmode\vadjust pre{\hypertarget{ref-hoffman_2014}{}}%
Hoffman, M. D., and Gelman, A. (2014). The no-u-turn sampler: Adaptively setting path lengths in hamiltonian monte carlo. \emph{Journal of Machine Learning Research}, \emph{15}, 1351--1381.

\leavevmode\vadjust pre{\hypertarget{ref-jarrow_2004}{}}%
Jarrow, R., and Protter, P. (2004). Structural versus reduced form models: A new information based perspective. \emph{Journal of Investment Management}, \emph{2}(2), 1--10.

\leavevmode\vadjust pre{\hypertarget{ref-kazianka_2016}{}}%
Kazianka, H. (2016). Objective bayesian estimation of the probability of default. \emph{Journal of the Royal Statistical Society: Series C: Applied Statistics}, 1--27.

\leavevmode\vadjust pre{\hypertarget{ref-nagl_2021}{}}%
Nagl, M., Havrylenko, Y., Pfeuffer, M., Jakob, K., Fischer, M., and Roesch, D. (2021). \emph{AssetCorr: Estimating asset correlations from default data}. Retrieved from: \url{https://CRAN.R-project.org/package=AssetCorr}

\leavevmode\vadjust pre{\hypertarget{ref-nishio_2019}{}}%
Nishio, M., and Arakawa, A. (2019). Performance of hamiltonian monte carlo and no-u-turn sampler for estimating genetic parameters and breeding values. \emph{Genetics Selection Evolution}, \emph{51}, 1--12.

\leavevmode\vadjust pre{\hypertarget{ref-park_2010}{}}%
Park, Y., Sirakaya, S., and Kim, T. (2010). A dynamic hierarchical bayesian model for the probability of default. \emph{Center for Statistics and the Social Science Working Paper}, \emph{98}.

\leavevmode\vadjust pre{\hypertarget{ref-pfeuffer_2020}{}}%
Pfeuffer, M., Nagl, M., Fischer, M., and Rösch, D. (2020). Parameter estimation, bias correction and uncertainty quantification in the vasicek credit portfolio model. \emph{Journal of Risk}, \emph{22}(4), 1--29.

\leavevmode\vadjust pre{\hypertarget{ref-tasche_2013}{}}%
Tasche, D. (2013). Bayesian estimation of probabilities of default for low default portfolios. \emph{Journal of Risk Management in Financial Institutions}, \emph{6}(3), 302--326.

\leavevmode\vadjust pre{\hypertarget{ref-vasicek_1987}{}}%
Vasicek, O. A. (1987). \emph{Probability of loss on loan portfolio}. KMV.

\leavevmode\vadjust pre{\hypertarget{ref-vasicek_2002}{}}%
Vasicek, O. A. (2002). The distribution of loan portfolio value. \emph{Risk}, \emph{15}(12), 160--162.

\end{CSLReferences}

\newpage

\hypertarget{appendix}{%
\section*{Appendix}\label{appendix}}
\addcontentsline{toc}{section}{Appendix}

\hypertarget{a1.-multifactor-credit-model}{%
\subsection*{A1. Multifactor credit model}\label{a1.-multifactor-credit-model}}
\addcontentsline{toc}{subsection}{A1. Multifactor credit model}

To generalize the model, we introduce multiple factors in order to model a credit portfolio through groups. We label the \(k\)-th credit of group \(j\) as \(I_k^{(j)}\) and, as before, we assume that they can be modeled through latent variables such that
\[
I_k^{(j)} = \pmb{1}_{\{X_k^{(j)}\leq u_j\}}.
\]

The latent variables are modeled through a multi-factor Gaussian copula, with the equations
\begin{equation}\label{eq:lat-var-gauss}
X_k^{(j)} = \sqrt{\rho_j} Z_j + \sqrt{1-\rho_j}\, \varepsilon_{k}^{(j)}
\end{equation}
where \(\pmb{Z} = (Z_1,...,Z_m)\) is a random vector of factors with distribution \(N_{(m)}(\pmb{0}, \Omega)\) and \(\varepsilon_k^{(j)}\) are i.i.d. normal standard r.v. representing idiosyncratic, independent of \(\pmb{Z}\). One can prove that, given the form in \eqref{eq:lat-var-gauss}, the vector of latent variables \(\pmb{X}\in \mathbb{R}^n\) can be written as
\[
\pmb{X} = R\pmb{Z} + Q\pmb{\varepsilon}
\]
with \(\pmb{\varepsilon}\) the vector of all idiosyncratic risks, and \(R\) and \(Q\) appropriate matrices. Thus, \(\pmb{X}\) follows a multivariate normal distribution with mean \(\pmb{0}\) and covariance matrix \(\Sigma = R\Omega R^T+ Q\) such that
\[
\Sigma_{hk} = \begin{cases}
1&\text{if}\quad i=j\\
\rho_i &\text{if}\quad h=h(i), \, k=k(i)\\
\rho_{ij} &\text{if}\quad h=h(i), \, k=k(j)
\end{cases}
\]
where \(k=k(j)\) indicates that the \(k\)-th latent variable belongs to group \(j\). Then, the bivariate vector \((X_h^{(i)}, X_k^{(j)})\) has a bivariate standard normal distribution with correlation \(\rho_{ij}\). We call \(\rho_i\) the intra-cohort correlation, and \(\rho_{ij}\) the inter-cohort correlation. The above construction defines the entries of \(\Omega\) such that
\[
(\Omega)_{ij} = \frac{\rho_{ij}}{\sqrt{\rho_i \rho_j}}.
\]

As before, conditional on the vector of factors \(\pmb{Z}\), we have that the defaults are independent and thus the total number of defaults for group \(j\), \(D_j\), as a binomial distribution with parameters \(N_j\) and probability \(\pi_j(Z_j)\) where
\[
\pi_j(Z_z) = \left( \frac{\Phi^{-1}(p_j)-\sqrt{\rho_j}Z_j}{\sqrt{1-\rho_j}} \right).
\]
Note that \(\pi_j\) only requires the knowledge of \(Z_j\) to be calculated, so actually \(D_j|Z_j \sim Bin(N_j,\pi_j(Z_j))\). However, the factors are established jointly through \(\Omega\).

\hypertarget{a2.-prior-sensitivity-analysis}{%
\subsection*{A2. Prior sensitivity analysis}\label{a2.-prior-sensitivity-analysis}}
\addcontentsline{toc}{subsection}{A2. Prior sensitivity analysis}

In this appendix, we perform prior sensitivity analysis the LH default series, which has low defaults but high default variance. The idea is to check how the different parameters that govern the priors behavior impact the posterior inferences. In particular, if the data can persuade the priors even with highly informative priors and a small sample size.

As mentioned before, the priors are given by \(f(\rho) = BetaP(0.5,\phi)\) and \(f(p|\rho) = BetaP(0.2,a\rho)\). We will keep the mean parameters fixed, and adjust the \(\phi\) and \(a\) parameters to observe how inferences are altered through a more informative prior. We use a sequence for both values such that \(\phi=1,5,10,...,50\) and \(a = 10,20,40,...,200\). The following figure shows both priors for each \(\phi\) parameter.

\begin{figure}[!ht]

{\centering \includegraphics{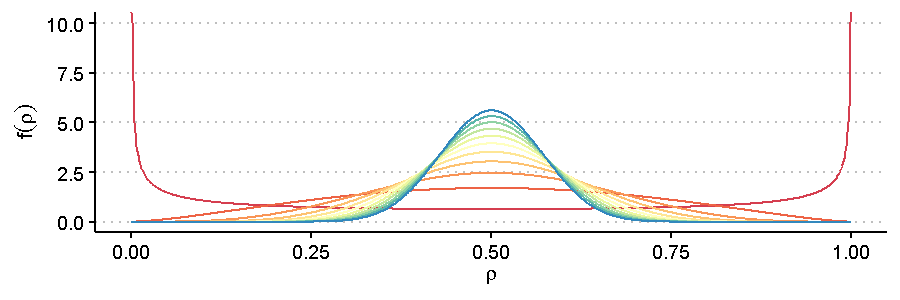} 

}

\caption{Prior distribution for $\rho$ for different values of $\phi$ and $a$}\label{fig:figA2-1}
\end{figure}

Figure \ref{fig:figA2-1} shows the prior distributions for \(\rho\) for the different parameters\footnote{Bluer colors represent higher values of both parameters.} of \(\phi\) and \(a\). We can observe how for very small values of \(\phi\) the distribution tends to extreme values, giving huge masses to 0 and 1. Such a behavior is undesirable, so we established before at 5. As we increase \(\phi\), the distribution concentrates more around 0.5.

For the default probability \(p\) we observe something similar, with a higher concentration the higher the values of both \(\phi\) and \(a\). In fact, the distribution is so concentrated that for the extreme values of \(\phi = 10\) and \(a = 100\), the simulated values for\footnote{Recall that \(f(p|\rho)\) is approximated through sampling of \(\rho\) and kernel smoothing.} \(p\) are at most 0.3, even though there is mass in the whole interval.

\begin{figure}[!ht]

{\centering \includegraphics{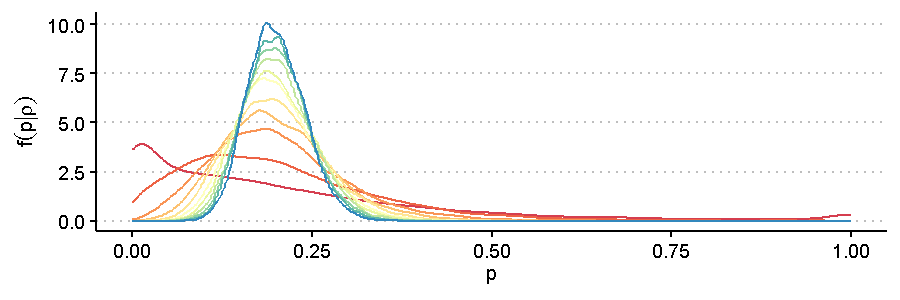} 

}

\caption{Prior distribution for $p$ for different values of $\phi$ and $a$}\label{fig:figA2-2}
\end{figure}

\begin{figure}[!ht]

{\centering \includegraphics{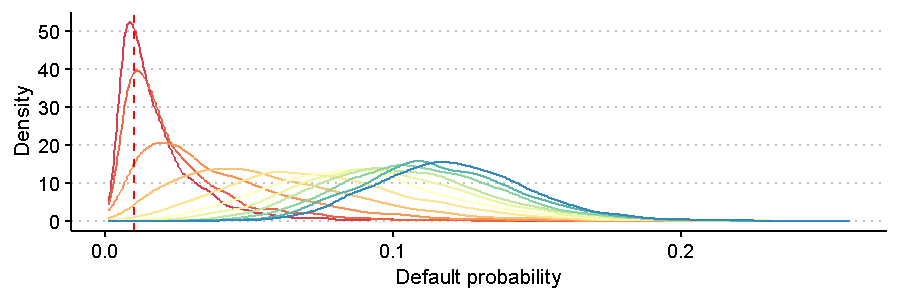} 

}

\caption{Posterior distribution for $p$ for different values of $\phi$ and $a$, with the real value in red}\label{fig:figA2-3}
\end{figure}

Given the samples, we plot the posterior densities for the parameters, for each of the different values of \(\phi\) and \(a\). Figure \ref{fig:figA2-3} shows the posterior kernel densities for the default probability. We observe that for small values of both \(\phi\) and \(a\), the posterior density is more precise, while with a higher right skewness. As both parameters increase, the distribution becomes flatter and gets closer to the prior mean of 0.2.

\begin{figure}[!ht]

{\centering \includegraphics{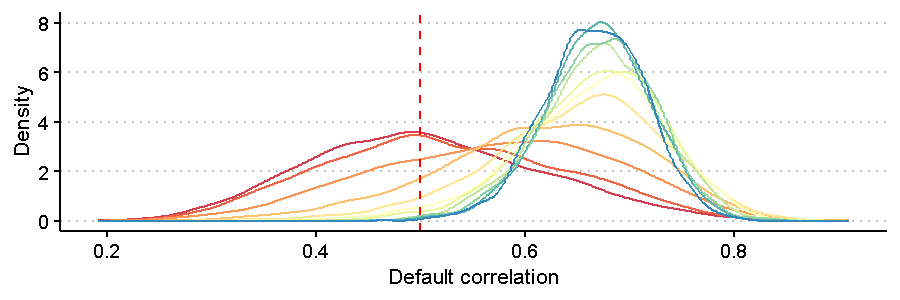} 

}

\caption{Posterior distribution for $\rho$ for different values of $\phi$ and $a$, with the real value in red}\label{fig:figA2-4}
\end{figure}

On the other hand, the behavior of the posterior distribution for the default correlation is not apparent, as seen in Figure \ref{fig:figA2-4}. As we increase the parameters, the distribution becomes more concentrated, but at higher values than the prior mean, closer to 0.7. Increasing the values of \(\phi\) will compensate and move again the posterior mean closer to 0.5, but it is clear that the behavior of \(p\) impacts the inferences on \(\rho\), as expected.

Thus, care must be taken in choosing appropriately weakly informative priors. Given a change in the prior distribution can lead to higher confidence on values further away from the true parameters. It is worth noting how, despite higher prior confidence on the ``true value'' of 0.5 through a more concentrated prior distribution for \(\rho\), the posterior distribution displaces to higher values with smaller variance.

\hypertarget{a3.-analysis-of-amle-method-for-low-correlation-default-series}{%
\subsection*{A3. Analysis of AMLE method for low correlation default series}\label{a3.-analysis-of-amle-method-for-low-correlation-default-series}}
\addcontentsline{toc}{subsection}{A3. Analysis of AMLE method for low correlation default series}

Given that for low correlation default time series, both intervals failed to capture the real value, we perform a detailed assessment of the estimates and bootstrap intervals given an increasing sample size. We start with a sample size of 3 and gradually increase it by 1 until we have 100 points, and observe how the estimates change as a function of the sample size. The following figure shows the results of the exercise.

\begin{figure}[!ht]

{\centering \includegraphics{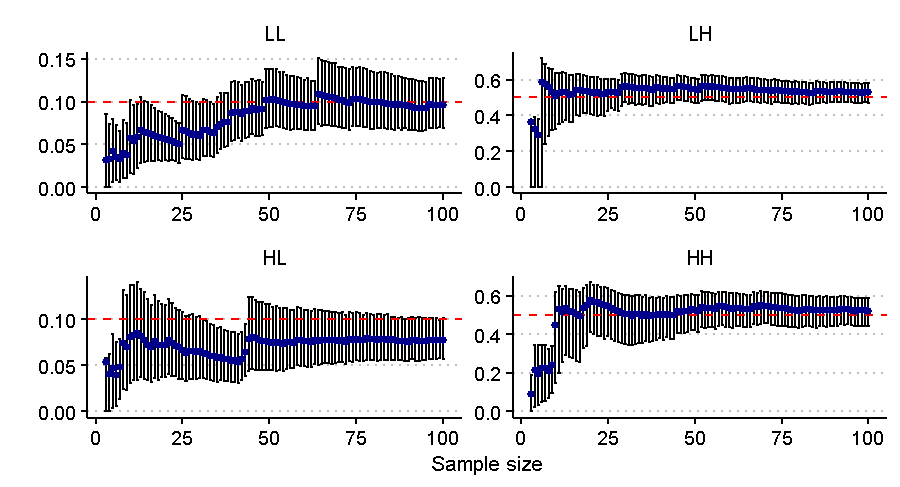} 

}

\caption{Defalt correlation 95\% bootstrap confidence interval, median value of the bootstrap replicates (blue) and real value for each type of series}\label{fig:figA3-1}
\end{figure}

We can observe how when the real correlation is high, the intervals tend to converge much faster to the real value of the correlation. In fact, for the LH series, with 6 data points the interval captures the real value. However, the low correlation estimates seem to struggle, with the HL series having an important bias even with 100 data points. Of course, taking up to 500 data points the intervals also converge for the HL series, but it is rare to have data of such lengthy default series.

This exercise shows the importance of correctly estimating the uncertainty of an estimate when we have few data points. When the correlation is high enough for the model and estimation method to capture, classical methods such as AMLE or CMM might be enough. However, in extreme cases, such as the high probability-low correlation default, it might be more appropriate to use, for example, Bayesian methods to better capture the uncertainty regarding our estimate. Indeed, as one can see in Figure \ref{fig:fig2-9}, the 95\% Bayes credible interval is much wider than both AMLE and CMM intervals for the low correlation series.

\end{document}